# Complex groundwater flow systems as traveling agent models

López-Corona O.[1,2], Padilla P.[3], Escolero O.[4], González T.[5] and Morales-Casique E.[4]

[1] Former: Posgrado en Ciencias de la Tierra, Instituto de Geología, Universidad Nacional Autónoma de México, Circuito Escolar, Cd. Universitaria México D.F.

[2] Currently: Theoretical Astrophysics, Instituto de Astronomía, Universidad Nacional Autónoma de México, Circuito Escolar, Cd. Universitaria México D.F.  olopez@astro.unam.mx.

[3] IIMAS, Universidad Nacional Autónoma de México, Circuito Escolar, Cd. Universitaria México D.F.

[4] Instituto de Geología, Universidad Nacional Autónoma de México, Circuito Escolar, Cd. Universitaria México D.F.

[5] Instituto de Geofísica, Universidad Nacional Autónoma de México, Circuito Escolar, Cd. Universitaria México D.F.

**Abstract**

Analyzing field data from pumping tests, we show that as with many other natural phenomena, groundwater flow exhibits a complex dynamics described by $1/f$ power spectrum. This result is theoretically studied within an agent perspective. Using a traveling agent model, we prove that this statistical behavior emerges when the medium is complex. Some heuristic reasoning is provided to justify both spatial and dynamic complexity, as the result of the superposition of an infinite number of stochastic processes. Even more, we show that this implies that non-Kolmogorovian probability is needed for its study, and provide a set of new partial differential equations for groundwater flow.

## 1   Introduction

Pink or 1/f noise (sometimes also called Flicker noise) is a signal or process with a frequency spectrum such that the power spectral density is inversely proportional to the frequency (Montroll and Shlesinger 1982; Downey 2001). This statistical behavior appears in such diverse phenomena as Quantum Mechanics (Bohigas et al. 1984; Faleiro et al. 2006; Haq et al. 1982; Relanyo et al. 2002), Biology (Cavagna et al. 2006; Buhl et al. 2006; Boyer and López-Corona 2009), Medicine (Goldberger 2002), Astronomy and many other fields (Press 1978). Recently the universality of 1/f noise has been related with manifestation of weak ergodicity breaking (Niemann et al. 2013) and with statistical phase transition (López-Corona et al. 2013)

In Geosciences the idea of self-organized criticality (SOC) associated with 1/f power spectrum showed to be important for example in modeling seismicity (Bak et al. 1987; Bak and Tang 1989; Bak and Chen 1991; Sornette et al. 1989). The basic idea of SOC is that large (spatially extended) interactive systems evolve towards a state in which a minor new event can have dramatic consequences. In seismicity this means that earthquakes contribute to organize the lithosphere both in space and time (Sornette et al. 1990). In this context, the lithosphere may be understood as an unstable and non-linear system made of hierarchy of interacting blocks and in which dynamics has a characteristic 1/f signal (Keilis-Botok 1990).

A particular active research field in Geoscience is the study of groundwater, which may be considered as a complex dynamic system characterized by non-stationary input (recharge), output (discharge), and response (groundwater levels). For example, groundwater levels in unconfined aquifers never reach steady state and may vary over multiple spatial and temporal scales showing fractal scaling characterized by inverse power law spectra (Zhang and Schilling, 2004 ; Jianting et al. 2012) . For a review on grounwater transport see Dentz et al. (2012).

Spectral analysis has proven to be a powerful analytical tool for the study of variations in hydrologic processes. Ever since Gelhar (1974) studied temporal variations of groundwater levels for the first time with spectral analysis, it has been widely used. Spectral densities have been used in the study of: the earth tides effect on water level fluctuations (Shih et al. 2000; Maréchal et al. 2002); temporal scaling in discharge (Tessier et al. 1996; Sauquet et al. 2008); scaling in hydraulic head and river base flow (Zhang and Schilling, 2004; Zhang and Li, 2005, 2006); water quality variations in space and time domains (Duffy and Gelhar 1985; Duffy and Al-Hassan 1988; Kirchner et al. 2000; Schilling et al. 2009b). Using the Detrended Fluctuation Analysis (DFA) method, Zhongwei and You-Kuan (2007) have proved that groundwater levels exhibit a $1/f$ behavior for large time scales.

The groundwater flow process may be considered as the motion of agents (water particles) in a heterogeneous medium (Traouez et al. 2001; Cortis and Knubdy 2006; Park et al. 2008). This problem is analogous to the model of traveling agents presented in (Boyer and López-Corona 2009). In that model, the agents are frugivorous animals who feed on randomly located vegetation patches, in a similar way to anomalously diffusing particles in a physical context. The displacement patterns of a variety of animals as albatrosses, bumble-bees, primates, gastropods, jackals, seals and sharks, among others (Viswanathan et al. 1999, Ramos-Fernández et al. 2004, Seuront et al. 2007, Atkinson et al. 2002, Austin et al. 2004, Sims et al. 2008) involve many spatio-temporal scales and are sometimes well described by Lévy walks. This is the case of the traveling agents of the model referred to (Boyer and López-Corona 2009). A good review on biological aspects of the subject may be found in Miramontes et al. (2012) and for Lévy process see Shlesinger et al. (1982),

Klafter et al. (1987) and Lomholt et al (2008),

The frequent occurrence of pink noise in a seemingly unrelated set of physical systems has prompted an extensive search for common underlying physical principles (Miller et al. 1993). In this paper we present a heuristic reasoning for the emergence of $1/f$ noise in groundwater and propose a new set of groundwater equations for flow in complex media (see electronic supplementary materials).

### 1.1 The traveling agent model

Let us consider a two-dimensional square domain of unit area with N fixed, point-like food patches randomly and uniformly distributed. Each patch contains an amount of food k.

Initially, an agent is located on a patch chosen at random. Then the following deterministic foraging rules are iteratively applied at every time step:

(i) The agent located at patch i feeds on that patch, the fruit content decreasing by one unit: $k_i \rightarrow k_i - 1$.

(ii) If ki has reached the value 0, the agent chooses another patch, j, such that $k_j / d_{ij}$ is maximal over all the allowed patches j ≠ i in the system, where $k_j$ is the food content of patch j and dij the Euclidean distance between patches i and j. With this rule, the next visited patch (the "best" patch) has large food content and/or is at a short distance from i. It was assumed that the travel from i to j takes one time unit.

(iii) The agent does not revisit previously visited patches.

This model produces complex trajectories that have been studied in detail in refs. (Boyer et al. 2006; Boyer et al. 2009) and discussed in connection with spider monkeys foraging patterns observed in the field (Ramos-Fernández et al. 2004). Most interesting is the fact that when this model is combined with a forest one, the coupled model exhibits

self-organized criticality and 1∕f power spectrum for biomass time series (Boyer and López-Corona 2009)

We propose that it is possible to use an equivalent model to study groundwater flow, conceptualizing it as the motion of water particles (agents) in a hydrogeological medium.

Assume the existence of a scale of support w where porous media properties can be measured. This scale of support is kept constant and is small enough such that, at the scale of the flow domain, w can be represented as a point-like quantity. Let us consider a two-dimensional square domain of unit area with N fixed, point-like Hydrogeological Units (HU) randomly and uniformly distributed. Each HU is characterized by its hydraulic flow potential, defined as $K_i = H_i / R_i$, where $H_i$ and $R_i$ are hydraulic head and hydraulic resistivity at point i, respectively. Thus $K_i$ has units of time.

Initially, an agent (water particle) is located on a HU chosen at random. Then the following deterministic motion rules are iteratively applied at every time step:

(i') An agent located in a HU stays there for a dimensionless time T proportional to $K_{max}/(K + a)$, where $K_{max}$ is the maximum hydraulic flow potential in the domain and a is an arbitrary normalization constant such that $K_{max} \gg a$. For $K \to 0$ then the waiting time is the maximum possible; For $K \to K_{max}$ then the waiting time is the minimum possible.

(ii') When an agent has spent T time in the $HU_i$, it chooses another $HU_j$, such that $\Delta K_{ij} / d_{ij}$ is maximal over all the allowed HU (i≠j) in the domain, where $\Delta K_{ij} / d_{ij}$ is the hydraulic flow potential difference between $HU_i$ and j, and $d_{ij}$ is the Euclidean distance between points i and j. With this rule, the next visited HU has the largest hydraulic flow potential gradient. It is assumed that the travel from i to j takes one time unit.

(iii') For a particular set of initial and boundary conditions, the agent does not revisit previously visited HU.

With this set of rules, both models, biological and groundwater flow, have the same statistical properties despite representing very different systems and then a direct analogy may be considered.

This traveling agent model exhibits some remarkable properties. Let us define the displacement of an agent R(t)=|R(t + t0) − R(t0)| with R(t) is the location of the agent at time t. For analysis, averages werte taken over different and independent realizations. If the hydraulic flow potential K follows an inverse power-law distribution $P(K) = cK^{-\beta}$, where c is an arbitrary constant and β is a coefficient that represents the medium homogeneity. When β is large (β>>1) the medium is very homogeneous, meaning that all HU have similar values of hydraulic flow potential. On the contrary when β is small (β~1) the medium is very heterogeneous, meaning that HUs with high hydraulic flow potential are numerous. The intermediate case is when β=3 and corresponds to a complex medium where HUs with high hydraulic flow potential are present but they are not so numerous.

## 2   Lévy walks and 1∕f dynamics

In recent works (Eliazar and Klafter 2009a, 2009b,2010) proved that Lévy walks and 1∕f are the result of systems which superimpose the transmission of many independent stochastic signals.

With this in mind, we proceeded to investigate if the power spectrum of the agent's motion follows a 1∕f dynamics. We found a non-trivial relationship between the homogeneity coefficient β, the motion of the traveling agent and the type of noise observed. These results (summarized in Table 1) are new and differ from previous work since now the motion of the agents is explicitly analyzed.

| Homogeneity | Medium type | Agent motion type | Displacement noise type |
|---|---|---|---|
| β=2 | Inhomogenous | Random Confined | White |
|  | Disordered |  | Uncorrelated |
| β=3 | Complex | Lévy | Pink (1/f) |
|  | Transition point | Fractal | Transition point |
| β=5 | Homogeneous | Brownian | Brown |
|  | Ordered |  | Highly correlated |

*Table 1   Relation between media homogeneity coefficient β, type of medium, agent motion, and the noise type observed.*

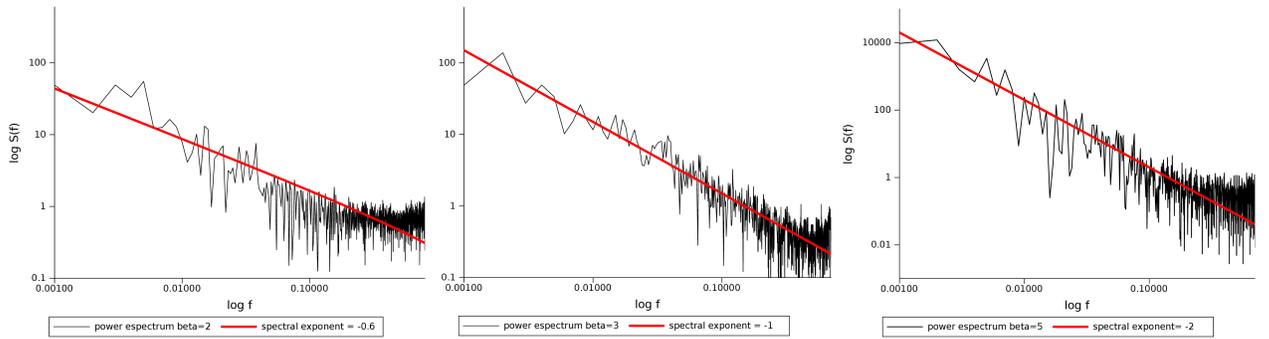

Figure 1   Power spectra for traveling agents with three values of homogeneity. First column β = 2, the medium is very inhomogeneous (disordered) and the signal is a white noise. Second column β = 3, the medium is complex and the signal is a pink noise. Third column 5, the medium is very homogeneous (ordered) and the signal is a brown noise. Power Spectrum is taken as $S(f) \equiv \tilde{R}(f)\tilde{R}(-f)$, where $\tilde{R}(f)$ is the Fourier transformation of the displacement calculated by a Fast Fourier Transformation technique.

Fifty time series for R(t) were generated using the implemented traveling agents model in (Boyer and López-Corona, 2009) which we propose is analogous to groundwater flow. Three values of β={2,3,5} were considered, corresponding to disordered, complex and ordered media. Then all the 50 power spectra were averaged and fitted by an inverse power law S(f) ~ f − λ. White noise correspond to a λ ≈ 0, pink to a λ ≈ 1, and brown to a λ ≈ 2.

These results show that the emergence of pink noise for a traveling agent in a heterogeneous medium depends on its degree of heterogeneity. Thus this dynamical behavior may naturally arise from the motion of agents in a complex medium. The agents may be frougivorous monkeys, and the complex medium a rain forest; or the agents may be water particles and the medium an aquifer with a complex geology. Our results suggest that 1/f noise may be a fingerprint of a statistical phase transition from randomness (low correlation associated with white noise), to predictability (high correlation associated to brown noise)   an idea suggested to us by Alejandro Frank (Personal communication, 2011) and discussed in (Fossion et al. 2010).

### 3   Study Case

As part of an academic collaboration between German Karlsruhe Institute of Technology (KIT) and Mexico's National University (UNAM), pumping tests were performed on a set of urban well in the metropolitan zone of the San Luis Potosi city in Mexico (ZMSLP), which hydrogeology is described in (Martinez et al. 2010, and Martinez et al. 2011).

The metropolitan area is located approximately 400 kilometers northwest of Mexico City. It lies in the San Luis Potosi valley in the west-centre of the state of the same name at an altitude between 1,850 and 1,900 meters above sea level. The area is flanked by the hills of Sierra San Miguelito to the west and Cerro San Pedro to the east; the hills have an altitude of more than 2300 meters. The climate is semi-arid with an average rainfall of 356 mm between 1989 and 2006, an average annual temperature of 17.68°C, and average annual potential evaporation of approximately 2,000 mm. The San Luis Potosi aquifer system underlies much of the surface endorheic basin. It consists of a shallow aquifer and a deep one, separated by a lens of fine material that permits very little interaction. The shallow aquifer is recharged by rainfall in the valley and the Sierra San Miguelito foothills, as well as by leaks from the urban water system. The deep aquifer is recharged in the Sierra San Miguelito and beyond. The 300 Km2 of shallow aquifer underlies the urban zone and its periphery. The thickness of the aquifer is within a range estimated at four to 60 meters, while the depth of the phreatic level has been reported in general terms at between five and 30 meters. The less deep levels are to be found within the urban zone and they deepen towards the east and northeast in the area of peripheral farmland, following the direction of the flow. The deep aquifer covers about 1980 Km2 and underlies the municipalities of San Luis Potosi and Soledad de G. Sanchez, as well as part of Cerro San Pedro, Mexquitic and Zaragoza. The aquifer consists of granular material and fractured volcanic rock, and is confined over most of the flat part of the basin. Usually, wells tapping this aquifer terminate at a depth of 350– 450 meters and exceptionally at 700 meters.

The time series from three pumping well tests, in the shallow aquifer, were analyzed. A pumping test is conducted to evaluate an aquifer by "stimulating" the aquifer through constant pumping, and observing the aquifer's response (drawdown) in observation wells. The power spectrum from all tests shows that there are two statistical regimes (Figure 2). The first regime is characterized by time periods from 101s to 103s and a $1/f$ noise statistical behavior, and the second one with periods of seconds or less and a white noise type of signal.

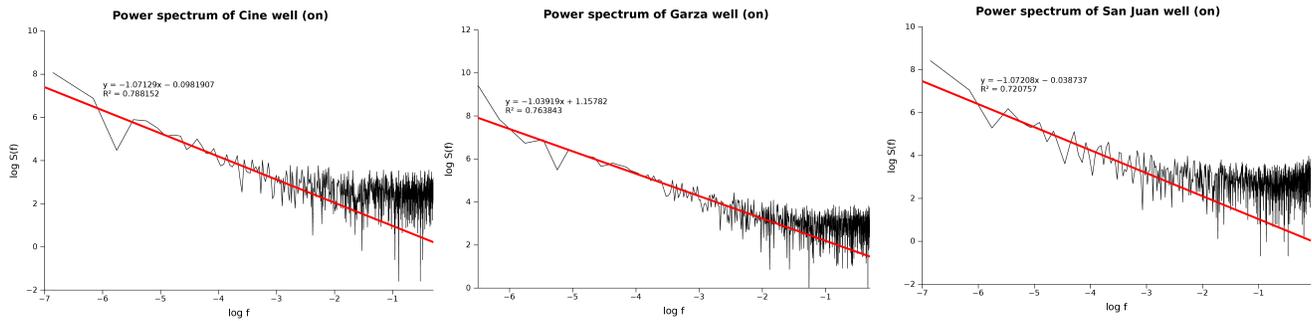

Figure 2   Power spectra for three pumping tests in the aquifer of San Luis Potosi City in Mexico. Drawdown data was acquired in a 3 seconds intervals basis, with a total of 1800 measurements. There are two statistical regimes 101s to 103s with a $1/f$ noise statistical behavior, and the second one with periods of seconds or less and a white noise type of signal.

## 4  Discussion and conclusions

Major sources of uncertainty have been identified in groundwater modeling. Model parameters are uncertain because they are usually measured at a few locations which are not enough to fully characterize the high degree of spatial variability at all length scales; thus, it is impossible to find a unique set of parameters to represent reality correctly. Stresses and boundary conditions are also uncertain; the extraction of water through wells and vertical recharge due to rain are not known exactly and they must be provided to the model; lateral boundaries are often virtual boundaries and water exchange through them is usually uncertain. Even model structure can be uncertain because a mathematical model is an approximation of reality and thus some physical processes are not completely known or partially represented (Neuman 2003). In fact, the problem of characterizing subsurface heterogeneity has been one of the biggest obstacles in constructing realistic models of groundwater flow (Fleckenstein et al. 2006). Koltermann and Gorelick (1996) and De Marsily et al. (1998) present a good review on the subject.

Prediction with classical deterministic process models is constrained by several mathematical limitations. For one side, there is measurement error, non-linearity and sensitivity to boundary conditions (chaos) and on the other side we most face model error and inaccessible or uncertain parameters and variables (Little and Bloomfield, 2010). For these reasons, systematic oversimplifications in groundwater problems have been commonly made, under the assumption that if the most important processes are identified, groundwater flow may be sufficiently characterized.

On the other hand, Kirchner and coworkers (2000) found that long-term, time series measurements of chloride, a natural passive tracer, in runoff in catchments exhibits a 1/f dynamics and later (Scher et al. 2002) gave a physical model to explain these founding in terms of CTRW.

Significant deviations from standard solutions have been observed in pumping tests (Raghanvan, 2004). Moreover, it has been reported that $1/f$ dynamics is observed in time series of pumping test (Zhongwei and You-Kuan 2007) and we showed evidence that support their findings. One approach to deal with this anomalous behavior has been to formulate the groundwater flow problem in the continuous time random walk (CTRW) framework (Cortis and Knubdy 2006). Alternatively we propose a traveling agent model for groundwater flow. The model proposed is an analogy of a previous one presented by (Boyer and López-Corona, 2009) which was used to construct time series for agent's mean-displacement. In agreement with field results, the model generates 1/f dynamics when the ambient where the agents move is complex. For this type of medium, the step length follows a power law distribution P(l)~l-a with a≈2 ; the waiting time distribution follows a power law y(t)~t−d with d=2 and the mean displacement a power law (R2)~Tg with g≈1.2 (Boyer and López-Corona, 2009; López-Corona, 2007). If the process was a CTRW then the following relationship should hold g=2+d−a and a value of g=2 would be expected (Klafter et al. 1995). This suggests that groundwater flow is even more complex than a CTRW, which in fact also occurs in spider monkeys foraging process for which g=1.7 (Ramos- Fernández et al. 2004). In this sense, the model proposed could be a forward step in the study of groundwater flow complexity.

Another advantage of the traveling agent model for explaining the emergence of $1/f$ is that we may identify in which type of hydrogeological medium this kind of dynamic behavior is observed. We proved that pink noise is present when the environment heterogeneities in which the agents are moving are distributed as a power law with a scaling exponent of $\beta = 3$, corresponding to a complex medium. At this respect, Labat (2011) has pointed out that the complex characteristics of karstic aquifers make their exploitation more complicated than other porous or fractured aquifers. These types of aquifers are spatially complex (as our $\beta=3$ medium) groundwater systems characterized by an inherent temporal non-stationarity and nonlinearity of their hydrological response.

(Eliazar and Klafter 2009a, 2009b) have proven that the 1/f statistical dynamics is originated by the superposition of an infinite number of stochastic processes. This suggests that for complex media (as karstic or rock fractured aquifer) no groundwater modeling simplification is valid. This ambient induces a $1/f$ noise and an infinite number of stochastic processes are in play. Therefore, the assumption that groundwater flow may be sufficiently characterized if the most important processes are identified is no longer valid.

Even more, the results may be interpreted also from a physical stand point; the observable macroscopic behavior of a hydrogeological system at a given location is the result of the superposition of different physical processes at different scales, such as: diurnal barometric variations that affect groundwater levels, temporal fluctuations in recharge rates, moon's gravitational effects over the aquifer, tide variations in coastal aquifers, variations in the income flow from rivers and discharge through base flow, temporal increase on total stress due to trains, the effect of extraordinary recharge events provoked by an hurricane presence, and the regime of operation of wells in the area. In Labat et. al. (2011) it has been proved, using DFA analysis, that in Karstic stream flow fluctuations there are three distinct temporal scale ranges: from 1 h to around 100 h, from around 100 h up to 1 year and scales larger to 1 year. Fluctuations in flow show a clearly anti-correlated behavior on time scales above 1 year, with a slope around 0.3 corresponding to white noise. In the intermediate regime from a few days up to 1 year, a positive Hurst effect is observed, with a slope around 0.8 (almost a 1/f noise) as expected. On time scales below the crossover at a few days, the scaling behavior is highly non-stationary and corresponds to a random walk with positively correlated steps (with a slope around 1.75, near a Brown noise type). The authors explain these findings from a hydrogeological point of view. The first temporal scale, 1 to 100 h, is interpreted as the rapid response of the aquifer (associated with the main drain in the karstic system) to the rainfall; the second temporal scale, 100 h to 1 year, is the global response of the aquifer to rainfall input including the temporal structure of the peak flow; the third temporal scale, larger than 1 year, corresponds to the annual response of rainfall input including the regulation of the discharge via annex systems in the saturated zone. It has also been suggested that an explanation for the scale invariance of groundwater levels involve the system response to constantly changing driving inputs and boundary conditions, including boundaries imposed by management regimes, (Little and Bloomfield, 2010). In this way, the 1/f power spectrum observed in groundwater time series may be originated by both, complexity of the geological medium and the presence of complex external factors (as time dependent boundary conditions).

Given this, either we accept that these types of complex groundwater systems are not suitable of being modeled or we learn to deal with this infinite superposition of stochastic processes. Once groundwater flow is modeled on a traveling agent framework, we propose to describe it as a spatially extended game. Using this approach we have been able to deduce a set of partial differential equations starting from the discrete description of the model (the details of the derivation are presented as Electronic Supplementary Material). The probability of finding an agent (water particle) in the position (x, y) at the time t obeys

$$\partial_t P(x, y, t) = \text{div}[e(x, y, t)\nabla P], \qquad (1)$$

where e(x, y, t) is the strategy (micro-physics of the flow process) that the agent in (x, y) plays at time t. The strategy in turn obeys the equation

$$\partial_t e(x, y, t) = -\text{div}[D_1(x, y, t)e(x, y, t)] + \nabla^2[D_2(x, y, t)e(x, y, t)]. \qquad (2)$$

While in continuous time random walk approaches few parameters suffice to describe a complex system, Eqs (1) and (2) introduce field (x,y) dependent diffusion and drift coefficients, and thus represent a quite complex approach. At this respect, Godec and Metzler (2013) has provied an exact expression for the diffusion coefficient in anomalous diffusion process modeled by Lévy walks under linear response regime.

If you take the simple case when $e(x, y, t)$ is a constant (considering that the porous medium es relative constant in the observation time scale, and it is sufficiently homogeneous and isotropic, which by the way are common assumptions in hydrogeology), then you recover the classical groundwater flow equation $S_s \partial h/\partial t = k\nabla^2 h$. Our equations then satisfy the correspondence principle since they recover classical formulation and establish the ground for new insights of groundwater flow process, other porous media transport phenomena and even in Game Theory.

Typically, a system is considered complex when it is constituted from a large number of subsystems that interact strong enough, but there is another source of complexity that has been widely ignored. A system is also complex when the systems itself changes over time in the same scale of its dynamics, which is be the case in some karstic aquifers. This second source of complexity is taken into account directly in our equations making a contribution in this respect and might have some important interpretation in Game Theory.

Finally, most interesting, using the traveling agent model described in the method section, we proposed (Lopez-Corona et al. 2013) that 1/f noise is a fingerprint of a statistical phase transition, form randomness (disorder – white noise) to predictability (order – brown noise).  In this context, one may interpret Labat (2011) results as follows: first temporal scale (from 1 to 100 h) represents a the rapid response of the aquifer and should be dominated by random processes (white noise); the second (100 h to 1 year) is the global response of the aquifer to rainfall input including the temporal structure of the peak flow one may be interpreted as a complex (with multiple spatio and temporal scales included) process (1/f noise); and as the third correspond to mean (1 year or more) response is a more predictable process (brown noise). We have then a transition from randomness to predictability consistent with power spectra exponent values. In this way, the results of

Labat et. Al. (2011) is only one example of a universal statistical kind of phase transition.

**Acknowledgements**

This work was supported by CONACyT scholarship within the graduate program in Earth Sciences at Universidad Nacional Autonoma de Mexico (UNAM). Fruitful discussions with Tomas González are gratefully acknowledged

# Electronic Supplementary Materials for Complex groundwater flow systems as a traveling agent models

López-Corona O. ,[1,2]*Padilla P.,[3] Escolero O. ,[4] González T.[5] and Morales E.[5]

April 29, 2014

[1]Former: Posgrado en Ciencias de la Tierra, Instituto de Geología, Universidad Nacional Autónoma de México, Circuito Escolar, Cd. Universitaria México D.F.
[2] Currently: Theoretical Astrophysics, Instituto de Astronomía, Universidad Nacional Autónoma de México, Circuito Escolar, Cd. Universitaria México D.F.
[3] IIMAS, Universidad Nacional Autónoma de México, Circuito Escolar, Cd. Universitaria México D.F.
[4]Instituto de Geología, Universidad Nacional Autónoma de México, Circuito Escolar, Cd. Universitaria México D.F.
[5] Instituto de Geofísica, Universidad Nacional Autónoma de México, Circuito Escolar, Cd. Universitaria México D.F.
*olopez@astro.unam.mx

## Derivation of the Flow Equations

Lets consider a group of agents moving in a square lattice according with a strategy $e$ function of actual location and neighbors density. The probability of finding an agent in an arbitrary node $(i\Delta x, j\Delta y)$ at the time $k\Delta t$, as in Figure 1, is

$$P(i\Delta x, j\Delta y, k\Delta t). \qquad (1)$$

The probability of an agent originally in $(i_0\Delta x, j_0\Delta y)$ at the time $k_0\Delta t$ walk to $(i\Delta x, j\Delta y)$ at the next time $k\Delta t = (k+1)\Delta t$ is

$$\begin{aligned}
P(i\Delta x, j\Delta y, k\Delta t) - P(i_0\Delta x, j_0\Delta y, k_0\Delta t) &= e^t_{Uij}\left[P(i\Delta x, (j+1)\Delta y, k\Delta t) - P(i_0\Delta x, j_0\Delta y, k_0\Delta t)\right] \\
&+ e^t_{Dij}\left[P(i\Delta x, (j-1)\Delta y, k\Delta t) - P(i_0\Delta x, j_0\Delta y, k_0\Delta t)\right] \\
&+ e^t_{Rij}\left[P((i+1)\Delta x, j\Delta y, k\Delta t) - P(i_0\Delta x, j_0\Delta y, k_0\Delta t)\right] \\
&+ e^t_{Lij}\left[P((i-1)\Delta x, j\Delta y, k\Delta t) - P(i_0\Delta x, j_0\Delta y, k_0\Delta t)\right]
\end{aligned} \qquad (2)$$

Where $e^t_{dij}$ is the strategy that the agent adopt based on the density neighbors difference between his actual position and the next one in $d$ direction, so

$$\begin{aligned}
e^t_{Uij} &= e\left(\left\|(\nabla P)^t_{i,j+1}\right\|\right) \\
e^t_{Dij} &= e\left(\left\|(\nabla P)^t_{i,j-1}\right\|\right) \\
e^t_{Rij} &= e\left(\left\|(\nabla P)^t_{i+1,j}\right\|\right) \\
e^t_{Lij} &= e\left(\left\|(\nabla P)^t_{i-1,j}\right\|\right)
\end{aligned} \qquad (3)$$

If we define $\delta P^t_{Uij} = P(i\Delta x, (j+1)\Delta y, k\Delta t) - P(i_0\Delta x, j_0\Delta y, k_0\Delta t)$, and so for the rest of directions, then Eq.2 became

$$P(i\Delta x, j\Delta y, k\Delta t) = P(i_0\Delta x, j_0\Delta y, k_0\Delta t) + e^t_{Uij}\delta P^t_{Uij} + e^t_{Dij}\delta P^t_{Dij} + e^t_{Rij}\delta P^t_{Rij} + e^t_{Lij}\delta P^t_{Lij}. \qquad (4)$$

Which is the discrete form of the anisotropic diffusion equation[1]

$$\frac{\partial P(x,y,t)}{\partial t} = div\left[e(\|\nabla P\|)\nabla P\right]. \qquad (5)$$

The discrete anisotropic diffusion equation 4 could be rewritten as[2]

$$P^{t+1}_a = P^t_a + \frac{\lambda}{|\eta_a|}\sum_{b\in\eta_a} e(\delta P_{a,b})\,\delta P_{a,b} \qquad (6)$$

|   | C | D |
|---|---|---|
| C | R | S |
| D | T | P |

Table 1: Canonical payoff matrix for classical Prisoner's dilemma. Where T stands for Temptation to defect, R for Reward for mutual cooperation, P for Punishment for mutual defection and S for Sucker's payoff. To be defined as prisoner's dilemma, the following inequalities must hold: $T > R > P > S$

with $a$ the actual position, $\eta_s$ the neighboring position of $a$, $|\eta_a|$ the number of first neighbors of the the agent in $a$ and $\lambda \in \mathbb{R}^+$ a constant that define the diffusion rate.

Of course, the functional form of $e^t_{dij}$ could be a more generic one as

$$e^t_{dij} = e^t_{dij}(x, y, t) \tag{7}$$

in which case the general anisotropic diffusion equation is obtained

$$\frac{\partial P(x,y,t)}{\partial t} = div\left[e(x,y,t)\nabla P\right] = \nabla e \cdot \nabla P + e(x,y,t)\nabla^2 P. \tag{8}$$

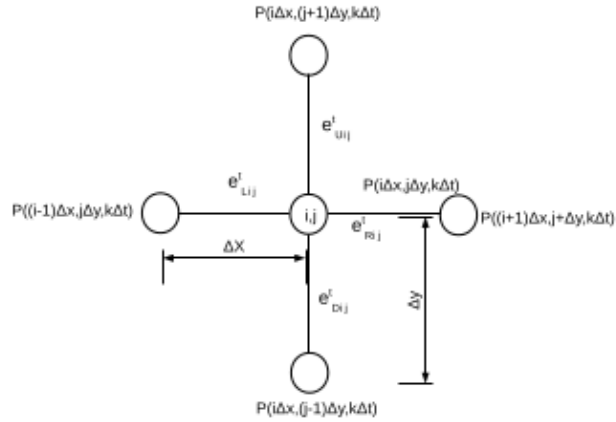

Figure 1: Diagram1

Even more, one could extend this analisis to a n players game and use the Evolutionary Theory of Nowak. In that context, if we relate $e$ with a particular game with payoff matrix $A$, and then $e^t_{hij}$ is the fraction of agents in the $(i,j)$ position that adopt strategy $h$ at time $t$. The corresponding replication equation is then

$$e^{k\Delta t}_{hij} - e^{k_0\Delta t}_{hij} = e^{k_0\Delta t}_{hij}(f_h - \phi_h) \tag{9}$$

where $f_h$, $\phi_h$ are the fitness and mean fitness of strategy $h$.

As an example, let be the matrix $A$ with elements $a_{hh'}$ the payoff of a Prisoner's dilemma game given in Table, then the corresponding fitness are

$$\begin{array}{rcl} h = C \Rightarrow f_C & = & (R)\, e^{k_0\Delta t}_{hij} + (S)\, e^{k_0\Delta t}_{hij} \\ h = D \Rightarrow f_D & = & (T)\, e^{k_0\Delta t}_{hij} + (P)\, e^{k_0\Delta t}_{hij} \end{array} \tag{10}$$

And the mean fitness is

$$\begin{aligned} \phi & = \sum_\gamma f_\gamma e^{k_0\Delta t}_{\gamma ij} e^{k_0\Delta t}_{\gamma(i+1)j} + \ldots \\ & + \sum_\gamma f_\gamma e^{k_0\Delta t}_{\gamma ij} e^{k_0\Delta t}_{\gamma(i-1)j} + \ldots \\ & + \sum_\gamma f_\gamma e^{k_0\Delta t}_{\gamma ij} e^{k_0\Delta t}_{\gamma i(j+1)} + \ldots \\ & + \sum_\gamma f_\gamma e^{k_0\Delta t}_{\gamma ij} e^{k_0\Delta t}_{\gamma i(j-1)} \end{aligned} \tag{11}$$

Taking into account equations 10 and 11 and that $e_{hij}^{k\Delta t} - e_{hij}^{k_0\Delta t}$ is the discrete form of the time derivative, the replicator equation can be rewritten [3] in matrix form as

$$\frac{dE}{dt} = [\Lambda(t), E(t)].\qquad(12)$$

Where $E, \Lambda$ are two matrix with elements

$$E_{hh'} = (e_h e_{h'})^{1/2}\qquad(13)$$

and

$$\Lambda_{hh'} = \frac{1}{2}\left[\left(\sum_{k=1}^{n} a_{hk} e_k\right)(e_h e_{h'})^{1/2} - (e_h e_{h'})^{1/2}\left(\sum_{k=1}^{n} a_{hk} e_k\right)\right],\qquad(14)$$

and the square brackets [] in equation 12, stands for commutation operation. It has been demonstrated that quantum game theory is a generalization of game theory an that the replicator equation 12 is equivalent to von Neumann equation

$$i\hbar \frac{d\rho}{dt} = [H, \rho]\qquad(15)$$

Where $\rho$ is the density matrix, a a self-adjoint (or Hermitian) positive-semidefinite matrix of trace one, that describes the statistical state of a quantum system. The $H$ operator is the hamiltonian of the system. The equivalence between the replicator and von Neumann equations are given by [3]

$$E \leftrightarrow \rho, \quad \Lambda \leftrightarrow -\frac{i}{\hbar} H.\qquad(16)$$

Via the master equation, it can be demonstrated [4, 5, 6] that the von Neumann equation leads to a Fokker-Planck equation of the form

$$\frac{\partial e(x,y,t)}{\partial t} = -div[D_1(x,y,t) e(x,y,t)] + \nabla[D_2(x,y,t) e(x,y,t)]\qquad(17)$$

where $D_1$ and $D_2$ are traditionally associated with drift and diffusion.

In this game theory context $D_1(x,y,t)$ is associated with the fitness $f(x,y,t)$ (Eq.10) and $D_2(x,y,t)$ with the mean fitness $\phi(x,y,t)$ (Eq.11).

The master equation is a first-order differential equation that describe the time evolution of the probability of the system to be in a particular set of states. Typically the master equation is given by

$$\frac{d\vec{P}}{dt} = A(t)\vec{P}\qquad(18)$$

where $\vec{P}$ is a column vector of the states $i$, and $A(t)$ is the matrix of connections. Many physical problems in classical, quantum mechanics and other sciences, can be expressed in terms of a master equation. Examples of these are the Lindblad equation in quantum mechanics and as we mention above, the Fokker–Planck equation which describes the time evolution of a continuous probability distribution. For more hydrologically aplications of the master equation the reader may refer to [7].

We can finally enunciate the discrete spatially extended game in continuum terms. The probability of finding an agent in the position $(x, y)$ at the time $t$ is given by

$$\frac{\partial P(x,y,t)}{\partial t} = div[e(x,y,t)\nabla P].\qquad(19)$$

Where $e(x,y,t)$ is the strategy that the player in $(x,y)$ plays at the time $t$ and that obeys the equation

$$\frac{\partial e(x,y,t)}{\partial t} = -div[D_1(x,y,t) e(x,y,t)] + \nabla^2[D_2(x,y,t) e(x,y,t)].\qquad(20)$$